# Nanoscale silicon sensor – guided new insights into early metabolic response of *Escherichia coli* to ampicillin

Yingtao Yu[1†], Victoria Nolan[1†], Zheqiang Xu[1], Allison Jones[2], George Alhoush[1], Zhen Zhang[1*], Sanna Koskiniemi[2*]

[1]Division of Solid-State Electronics, Department of Electrical Engineering, Ångström Laboratory, Uppsala University, SE-75103, Uppsala, Sweden

[2]Department of Cellular and Molecular Biology, Uppsala University, SE-75124, Uppsala, Sweden

*Corresponding authors:
Prof. Zhen Zhang
Email: zhen.zhang@angstrom.uu.se
Prof. Sanna Koskiniemi
Email: sanna.koskiniemi@icm.uu.se

**Abstract:**

Antibiotic killing is often attributed to inhibition of specific cellular targets, yet metabolic processes can strongly influence drug efficacy. However, the relationship between metabolic responses and antibiotic lethality remains incompletely understood. Here, we employed silicon nanowire field-effect transistor (SiNWFET) sensors to monitor real-time metabolic responses of *Escherichia coli* to ampicillin (AMP). AMP treatment induced a biphasic extracellular pH signature, characterized by rapid acidification followed by alkalization. Metabolomic analyses revealed that the initial acidification resulted from organic acid secretion, whereas the subsequent alkalization was associated with altered amino acid metabolism and formate flux. Using metabolic and respiratory mutants, we found that these extracellular signatures reflected pathway-specific metabolic rewiring that predicted bacterial killing. AMP lethality strongly correlated with ATP dynamics and formate secretion: strains exhibiting larger AMP-induced ATP increases and greater formate secretion showed enhanced susceptibility. In contrast to literature, changes in NADH and NADPH levels did not support redox stress as the primary bacterial-killing mechanism. Together, our findings identify formate metabolism as a key pathway underlying the elevated ATP levels associated with AMP bactericidal activity. These results demonstrate that SiNWFET sensors provide a versatile label-free tool for probing antibiotic mechanisms, rapidly assessing bacterial susceptibility, and potentially guiding antimicrobial therapy development.

## Introduction

Over the past few decades, tremendous efforts have been dedicated to the fundamental understanding of antibiotics and their mechanisms of action. Nevertheless, those studies have primarily been based on profiling their direct target cellular processes, such as DNA replication, protein synthesis, cell wall synthesis, *etc.* [1], [2], [3], [4], resulting in a somewhat oversimplified view that antibiotic-induced cell death exclusively arises from the inhibition of specific targets [5]. An overlooked fact is that these processes play critical roles in cell growth and consume substantial portions of the metabolic output. Therefore, it is possible that cellular metabolism could be perturbed as a downstream reaction to antibiotic-target interactions. Recent evidence suggests that bactericidal antibiotics can accelerate cellular respiration and trigger metabolic overflow, causing accumulation of reactive oxygen species resulting in oxidative damage[6], [7], [8], [9]. In contrast, bacteriostatic antibiotics can suppress metabolism and thus effectively keep bacteria in the stationary phase of growth [10]. Furthermore, it has been demonstrated that active respiration potentiates bactericidal antibiotic lethality, whereas metabolic repression by bacteriostatic antibiotics attenuates bactericidal activity of antibiotics [10]. Therefore, a deeper exploration into the role of metabolism is crucial for understanding antibiotic efficacy. This can further facilitate the development of new treatment strategies.

Active bacterial metabolism, such as glycolysis, the TCA cycle, amino acid metabolism, *etc*, results in the production of metabolic byproducts which are secreted into the external environment. This causes changes in extracellular pH that can serve as important indicators of bacterial metabolic activity. In our previous work, we successfully used silicon nanowire field-effect transistors (SiNWFETs) to real-time profile bacterial metabolism, via monitoring the metabolism-induced acidification in the growth media[11], [12]. This was used as a basis for rapid antibiotic susceptibility testing (AST), providing results within 30 min. Compared with traditional fluorescent detection, our SiNWFET is an all-electrical, label-free, nondestructive, real-time method with multiplexing capability. But whether the SiNWFET could also detect changes to metabolism mediated by specific antibiotics or their bacteriocidality was not clear.

In this work, we used the SiNWFET sensors to reveal the early metabolic response of *Escherichia coli* (*E. coli*) to the bactericidal antibiotic ampicillin (AMP). AMP triggered a striking biphasic extracellular pH signature; rapid acidification followed by alkalization, driven by organic acid secretion and subsequent shifts in amino acid and formate metabolism. By engineering mutants targeting central metabolic pathways, we show that these metabolic rewiring events directly shape both the pH signatures and antibiotic lethality. In particular, AMP-induced killing correlated with media alkalization and formate secretion, as well as ATP dynamics, while redox changes did not correlate with cell death as proposed in the redox stress killing theory. Together, these findings uncover predictive metabolic markers of AMP efficacy and emphasize the power of SiNWFET sensors to link real-time extracellular signals to intracellular pathways and antibiotic outcomes.

## Results

### Metabolic responses to AMP exposure dictate pH changes in culture supernatant

We previously demonstrated SiNWFET devices with multiplexed readout system capable of rapid antibiotic susceptibility test on chip (Fig. 1A) [11]. The SiNWFET has an n-type 100-nm channel (Fig. 1B) and exhibits a subthreshold swing of ~110 mV/dec with an on-to-off current ratio ($I_{on}/I_{off}$) of ~$10^6$ (Fig. S1A). The change in threshold voltage, $\Delta V_T$ vs. time curve of a SiNWFET sensor measured in different pH buffer solutions is shown in Fig. S1B. The threshold voltage of the SiNWFET shifts positively with decreasing pH, showing a near-Nernstian response with a sensitivity of 56.0 ± 1.2 mV/pH (Fig. S1B). Under optimized experimental conditions, *E. coli* MG1655 (hereafter referred to as WT throughout the paper) acidified the media and shifted the $V_T$ to about 60 mV (corresponding to pH 6.2) within 25 min in LB (Fig. 1C). The $\Delta V_T$ signal was significantly attenuated by AMP concentrations of 50 mg/L or higher. The higher concentrations of AMP (1 g/L and 100 mg/L) show not only the halting of media acidification but a reversal after 25 min, consistent with our previous work (Fig. 1C) [11].

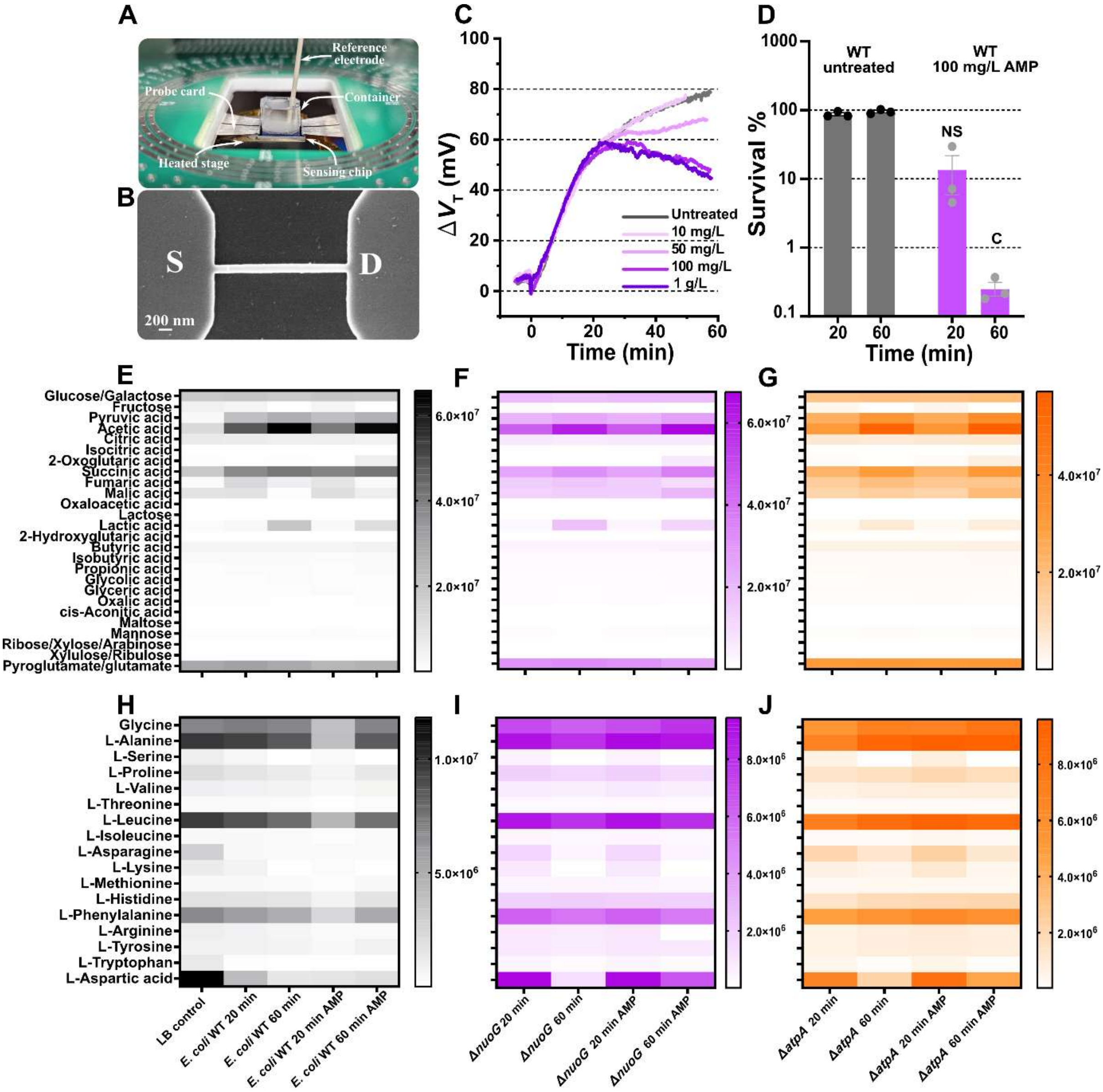

**Figure 1. On-chip electrical measurements using our SiNWFET devices and analysis of metabolites profiled during the electrical measurements. A)** Image of the on-chip multiplexing test system. B) SEM image of a single SiNWFET device. **C)** Detected $\Delta V_{th}$ signal from *E. coli* WT (OD = 6) in LB at 37°C under different AMP conditions. **D)** Survival % of WT exposed to 100 mg/L AMP or untreated at 20 or 60 min. Statistical significance was determined against AMP treated WT at the same time point using lognormal Welch's t-test. (NS = not significant, A = P < 0.05, B = P < 0.01, C = P < 0.001 and D = P < 0.0001). **E-I)** Metabolomics analysis of organic (**E-G**) and amino acids (**H-J**) secreted by WT (**E, H**), Δ*nuoG* (**F, I**) and Δ*atpA* mutants (**G, J**) with and without AMP.

To confirm that the antibiotics were indeed arresting cell growth, we performed time-kill experiments in parallel. WT subjected to 100 mg/L showed a rapid decline in cell viability under the device conditions, with 1-log of killing after 20 min and 2-3 logs after 60 minutes (Fig. 1D).

To further understand what contributes to media acidification observed during the on-chip tests,

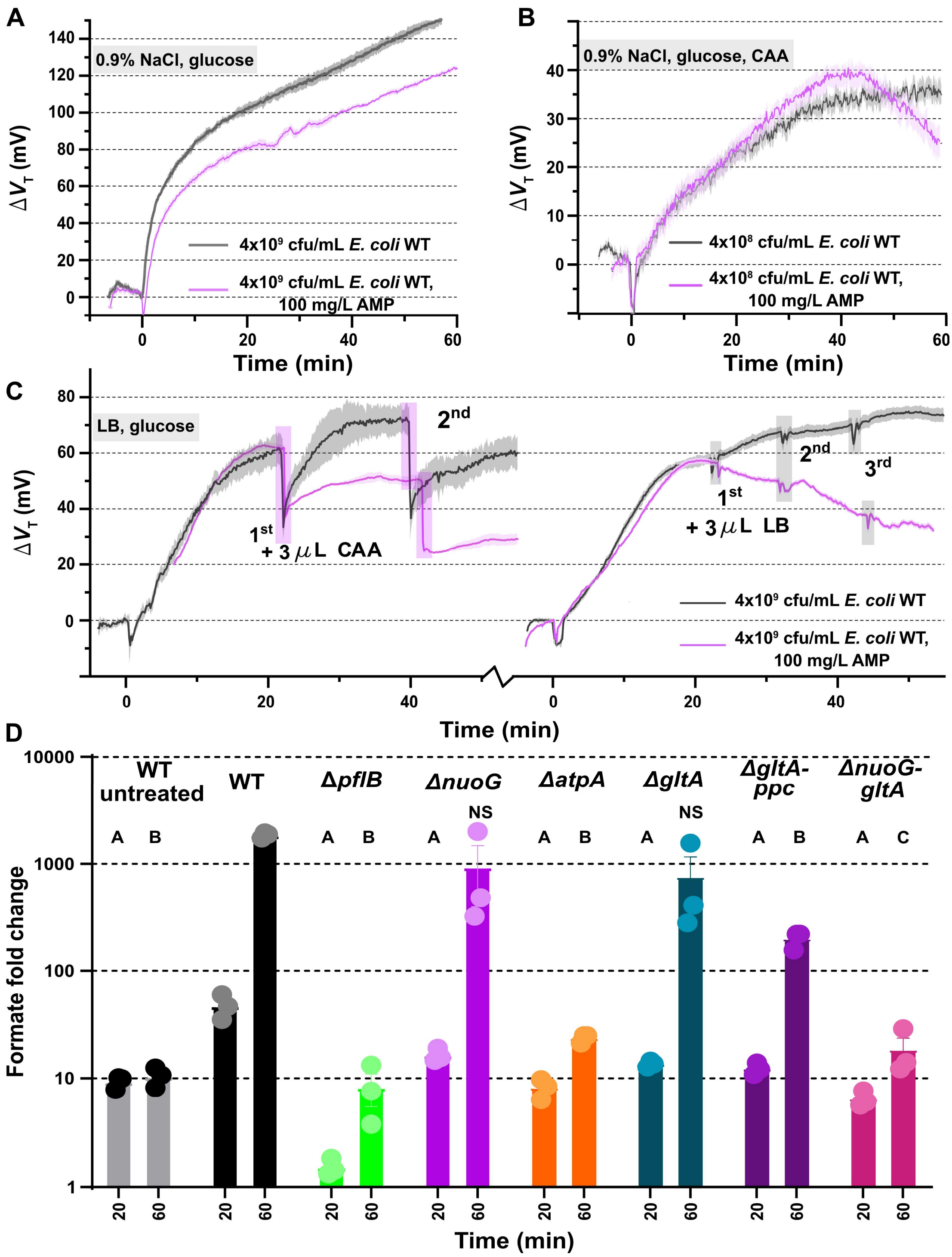

**Figure 2. Amino acid and formate metabolism contribute to perturbations in media acidification/alkalization. A-B)** $\Delta V_T$ vs. t curve of WT in 0.9 % NaCl with 1% glucose, with/without AMP **(A)** and supplemented with 4 g/L CAA **(B)**. **C)** $\Delta V_T$ vs. t curve of WT in LB supplemented with 3 µL CAA (100 g/L) (left) or 3 µL LB (right). **D)** Fold changes in secreted formate of AMP treated WT and select metabolic mutants *(ΔpflB, ΔnuoG, ΔatpA, ΔgltA, ΔgltA-ppc, ΔnuoG-gltA*). Statistical significance was determined against AMP treated WT at the same time point using lognormal Welch's t-test. (NS = not significant, A = $P < 0.05$, B = $P < 0.01$, C = $P < 0.001$ and D = $P < 0.0001$).

metabolomic analysis was performed on supernatant media collected during the on-chip tests and analyzed for organic acid metabolites and amino acids. Media acidification was primarily driven by acetic acid, succinic acid, pyruvic acid and lactic acid, regardless of AMP treatment (Fig. 1E). In response to AMP treatment, an increase in 2-oxoglutarate after 60 mins was seen as well as a decrease in most detected amino acids by 20 mins, followed by their re-appearance at 60 mins (Fig. 1H). This is consistent with the reversal of pH that was observed for WT in LB, where a decrease in pH was seen up to 20 mins followed by media alkalization at later time-points (Fig 1C).

**Media alkalization results from amino acid uptake and formate secretion**

The media alkalization could result from several factors, including: i) cell lysis in response to AMP, ii) uptake of acidic metabolites, or iii) secretion of basic metabolites. To investigate if alkalization resulted from cell lysis, we measured $OD_{600}$ values of the AMP-treated cultures to look for signs of lysis during this time frame. Only at the highest AMP concentration (1 g/L) could we observe a slight decrease in $OD_{600}$ values after 20 mins (Fig. S2A). At lower concentrations (100 mg/L and 50 mg/L AMP) $OD_{600}$ values continued to increase up to 60 mins. This indicates the pH increase observed at 20 mins is likely not due to cell lysis for 100 mg/L AMP. To further explore the impact of cell lysis on the media alkalization, the equivalent of 150 µL fully lysed cells (7.5 µL lysed WT from a concentrated overnight culture) was added to an on-chip WT culture after 60 min of incubation in four independent trials. No detectable increase in pH was observed (Fig. S2B). These results clearly indicate that the ~40 mV $\Delta V_t$ decrease observed after 1-h AMP treatment is not a consequence of cell lysis, but rather arises from metabolic responses induced by AMP treatment.

Because no uptake of acidic metabolites was observed following AMP treatment, we investigated whether amino acid secretion and uptake could instead provide buffering capacity and thereby contribute to the observed medium alkalinization. To test this, metabolic measurements were repeated on-chip in 0.9% NaCl supplemented with 1% glucose (Fig. 2A). Under these conditions, the pH response was markedly faster, with a $\Delta V_T$ of 109.3 mV reached within 25 min, compared with LB medium. This finding suggests that LB provides buffering capacity, potentially through its amino acid content. Although a ~20 mV difference between AMP-treated and untreated cells was still observed after 20 min in 0.9% NaCl, no medium alkalinization occurred. To assess the role of amino acids directly, 0.9% NaCl medium was supplemented with 4 g/L casamino acids (CAA) and analyzed on chip (Fig. 2B). In the presence of CAA, the pH response resembled that observed in LB, including the onset of medium alkalinization after approximately 20 min. This suggests that amino acid secretion may contribute to the alkalinization induced by AMP treatment. To further validate this hypothesis, 3 µL of a 100 g/L CAA solution was injected into 150 µL of on-chip LB medium during metabolic monitoring, mimicking amino acid secretion at 20 and 40 min after AMP treatment (Fig. 2C, left). CAA addition immediately induced a decrease in $V_T$ of approximately 20 mV, demonstrating that amino acids can directly contribute to medium alkalinization. In contrast, injection of the same volume of LB medium produced only a transient fluctuation attributable to the injection procedure itself, ruling out dilution or mechanical disturbance as the primary cause of the $V_T$ response (Fig. 2C, right). Although the amino acid concentration of LB is not precisely defined, it is substantially lower than that of the injected CAA solution based on its protein hydrolysate content (10 g/L tryptone and 5 g/L yeast extract compared to 100 g/L CAA). This may explain why addition of a small volume of LB did not measurably alter the medium buffering capacity. Taken together, these results suggest that amino acid secretion occurring between 20 and 60 min after AMP exposure contributes to the observed medium alkalinization during this period. However, amino acid secretion alone cannot explain why the AMP-treated samples do not exhibit a more acidic pH than the untreated controls at 20 min, when the amino acid concentration in the supernatant is decreased.

We next investigated whether the observed medium alkalinization could be explained by changes in formate metabolism, which was not included in the metabolomics analysis. Depending on pH, formate exists as either formic acid or formate and can therefore contribute to acid-base balance. In *E. coli*, formate transport is mediated by the FocA channel and regulated by the gatekeeper protein PflB (Fig. 3A). Under normal conditions, formic acid is exported at pH values above 6.8 and imported at pH values below 6.8 [13]. As the culture supernatant reached a pH of approximately 6 after 20 min, any additional formate secretion under these conditions would be expected to contribute to medium alkalization.

Measurements of the culture supernatants showed that formate concentrations increased approximately 50- and >1000-fold after 20 and 60 min of AMP exposure respectively (Fig. 2D). To assess the contribution of formate secretion to medium pH, we analyzed a Δ*pflB* mutant, which is impaired in formate secretion. In contrast to the WT, the Δ*pflB* mutant did not exhibit medium alkalinization after 60 min of AMP treatment (Fig. 3B), suggesting that formate secretion contributes to the pH increase. Consistent with this interpretation, direct formate measurements showed that the Δ*pflB* mutant secreted only minimal amounts of formate (Fig. S3). AMP treatment increased formate levels only ~8-fold after 60 min, similar to the untreated control (~10-fold), whereas WT cells displayed a dramatic induction of formate secretion in response to AMP (Fig. 2D). Interestingly, WT cells showed elevated formate secretion already after 20 min of AMP exposure. This early increase in formate may counteract the acidification expected from amino acid uptake and could therefore explain why AMP-treated cultures do not exhibit a lower pH than untreated controls at this time point. Taken together, these results indicate that AMP treatment alters both amino acid and formate metabolism, leading to changes in uptake and secretion patterns that influence extracellular pH. These metabolic adaptations may, in turn, affect AMP activity.

**Media alkalization is mainly driven by formate secretion in mutants affecting respiration.**

In addition to amino acid secretion, our results indicate that formate secretion is important to shift from medium acidification to alkalinization. Formate transport through the FocA channel occurs via $H^+$ symport and is regulated by intracellular and periplasmic pH, linking formate flux tightly to respiratory activity. Because AMP has previously been shown to accelerate respiration in *E. coli* [15], we investigated whether respiratory changes could influence formate secretion and thereby affect extracellular pH. To address this, two mutants impaired in the electron transport chain were analyzed: Δ*nuoG* and Δ*atpA* (Fig. 3A). *nuoG* encodes a subunit of NADH dehydrogenase I (complex I), whereas *atpA* encodes the α-subunit of ATP synthase, which couples the proton motive force to ATP synthesis. To establish the baseline effects of these mutations on extracellular acidification, $V_T$ was monitored in untreated cultures. Although both mutants showed increasing $V_T$ over time, their acidification rates were slightly lower than that of the WT (Fig. S4), indicating altered metabolism. Following AMP treatment, the Δ*nuoG* mutant exhibited a reversal in $\Delta V_T$ after 20 min, indicative of medium alkalinization, whereas the Δ*atpA* mutant showed little or no alkalinization (Fig. 3B). Analysis of the culture supernatants revealed that both mutants responded metabolically to AMP in a manner similar to the WT, including comparable organic acid production and increased levels of 2-oxoglutarate and malate after 60 min (Fig. 1F,G). In contrast to the WT, however, neither mutant showed significant changes in extracellular amino acid levels following AMP treatment (Fig. 1I,J). Despite this shared lack of amino acid secretion, the two mutants displayed markedly different alkalinization phenotypes, suggesting that amino acids alone cannot explain the observed pH responses. Formate measurements provided a stronger correlation with the alkalinization phenotype. Neither mutant showed increased formate secretion at 20 min after AMP treatment. However, after 60 min, the Δ*nuoG* mutant exhibited a ~1000-fold increase in extracellular formate, similar to the WT, whereas the Δ*atpA* mutant displayed only a modest ~20-fold increase (Fig. 2D). Importantly, the strain capable of strong formate secretion (Δ*nuoG*) also exhibited medium alkalinization, whereas the strain with limited formate secretion (Δ*atpA*) did not. Together with the results obtained for the Δ*pflB* mutant, these findings identify formate secretion as the primary determinant of AMP-induced medium alkalinization. While amino acid secretion likely contributes to buffering and may modulate the early pH response in the WT, the mutant analysis demonstrates that formate levels correlate more closely with the alkalinization phenotype and are therefore the dominant factor driving the observed pH changes.

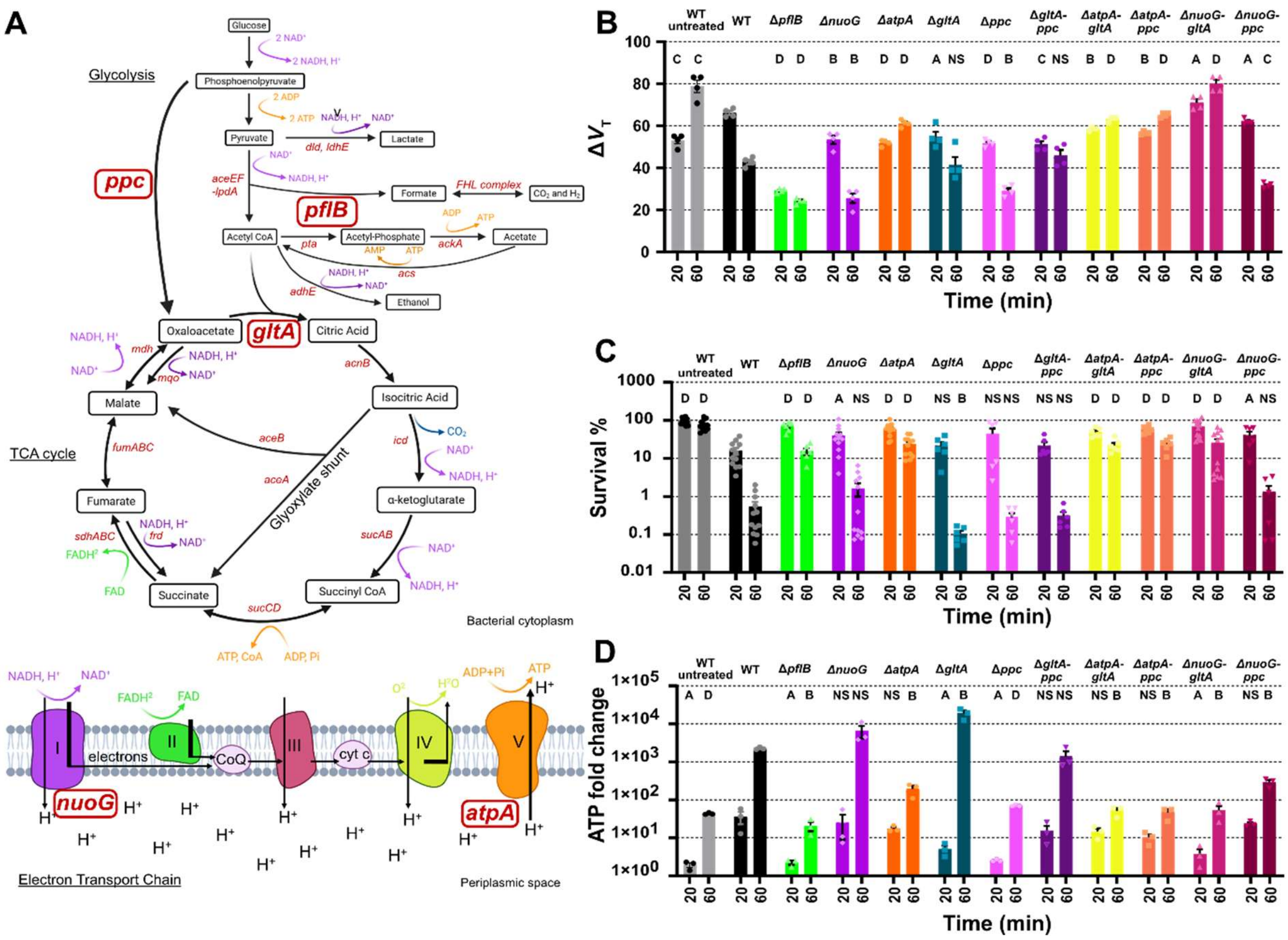

**Figure 3. Alterations to key points in central carbon metabolism result in perturbations to AMP treatment. A)** Schematic of the central carbon metabolism and the genes selected for removal (made in part by BioRender). **B)** $\Delta V_T$ changes observed for AMP-treated mutants at 20 min and 60 min (Time-series data can be found in Fig. S4). Welch's t-test identified significant differences in AMP-treated mutants compared to AMP-treated *E. coli* WT. **C)** Survival % of mutants exposed to AMP treatment at 20 min and 60 min. **D)** Fold changes in ATP in response to AMP treatment for *E. coli* WT and mutants.

Statistical significance was determined against AMP-treated WT at the same time point using Welch's t-test for B) and lognormal Welch's t-test for **C** and **D**. (NS = not significant, A = P < 0.05, B = P < 0.01, C = P < 0.001 and D = P < 0.0001).

**The metabolic switch to formate secretion is important for AMP bacteriocidality**

To determine whether medium alkalinization was linked to AMP bactericidal activity, we performed time-kill experiments using mutants that differed in their alkalinization response. AMP susceptibility correlated closely with the presence of medium alkalinization. Following 60 min of AMP treatment, only ~0.1% of WT and Δ*nuoG* cells survived, and both strains exhibited clear medium alkalinization. In contrast, Δ*atpA* and Δ*pflB*, which showed little or no alkalinization, displayed substantially higher survival (~10%) (Fig. 3C). These findings suggest a close association between AMP-induced alkalinization and bactericidal activity.

To further explore the relationship between metabolism and AMP susceptibility, metabolomic analyses were performed in 0.9% NaCl supplemented with glucose using WT, Δ*nuoG*, and Δ*atpA* strains. Under these conditions, the metabolic differences between the strains were more pronounced than in LB medium (Fig. S5A-F). Analysis of extracellular metabolites showed that the Δ*nuoG* mutant accumulated higher levels of fumarate and malate than WT (Fig. S5B), suggesting increased flux through either the glyoxylate shunt or the *ppc*-dependent reductive branch of the TCA cycle. In contrast, the Δ*atpA* mutant secreted very low levels of TCA cycle intermediates but accumulated pyruvate (Fig. S5C), consistent with its impaired formate secretion.

To investigate whether these differences in central carbon metabolism influenced AMP susceptibility, we generated mutants lacking genes that provide entry into the TCA cycle, either individually or in combination. The complete list of strains and their predicted metabolic consequences is provided in Table S3. In the WT background, deletion of *gltA*, *ppc*, or both genes resulted in modest reductions in medium acidification during the first 20 min compared with WT (Fig. S4B). However, their responses to AMP differed. The Δ*ppc* mutant displayed medium alkalinization similar to WT, whereas the Δ*gltA* and Δ*gltA-ppc* mutants showed weaker alkalinization. Consistent with this phenotype, Δ*gltA* and Δ*gltA-ppc* exhibited intermediate increases in extracellular formate (~200-fold; Fig. 2D). Despite these differences, all three strains remained highly susceptible to AMP, with survival rates of only ~0.1-0.3% after 60 min (Fig. 3C). Thus, although perturbation of TCA cycle entry affects formate secretion and extracellular pH, these effects are insufficient to substantially reduce AMP-mediated killing. In the Δ*atpA* background, deletion of *ppc* and/or *gltA* had no further effect on either medium alkalinization (Fig. S4C, Fig. 3B) or survival following AMP treatment (Fig. 3C). Similarly, the Δ*nuoG-ppc* double mutant closely resembled the Δ*nuoG* single mutant, displaying both medium alkalinization and high AMP susceptibility (~0.1% survival; Fig. 3B,C). In contrast, the Δ*nuoG-gltA* mutant showed a markedly different response. This strain failed to alkalinize the medium, resembling the Δ*atpA* and Δ*pflB* mutants, and exhibited increased survival (>10% after 60 min of AMP treatment) (Fig. 3B,C). Notably, this phenotype occurred despite a measurable increase in extracellular formate (~18-fold; Fig. 2D), indicating that a modest increase in formate is not sufficient to drive either strong alkalinization or full bactericidal activity. Taken together, these results strengthen the link between extracellular alkalinization and AMP-mediated killing. Across the mutant panel, strains exhibiting strong formate accumulation and alkalinization remained highly susceptible to AMP, whereas strains lacking alkalinization displayed substantially improved survival. These findings support a model in which formate flux, and the resulting changes in extracellular pH, are closely associated with the bactericidal activity of AMP.

**Increased ATP levels but not redox status indicate bacteriocidality**

Because disruption of the ETC and TCA cycle is expected to alter both ATP production and cellular redox balance, two processes previously implicated in AMP bactericidal activity, we next investigated whether these parameters could explain the differences in survival observed among the mutants. To assess the impact of AMP on energy metabolism, intracellular ATP levels were measured before and after AMP exposure. Most strains showed increased ATP levels following AMP treatment (Fig. 3D). However, the magnitude of this increase differed substantially between strains. Mutants with increased AMP survival (>10%), including Δ*atpA* (alone or in combination), Δ*pflB*, and Δ*nuoG-gltA*, displayed relatively modest increases in ATP levels after 60 min of treatment (~20- to 200-fold), comparable to untreated WT cells (~40-fold increase) (Fig. 3D). In contrast, WT cells exposed to AMP exhibited a >2000-fold increase in ATP levels. Similarly, several AMP-susceptible strains (<1% survival), including Δ*nuoG*, Δ*gltA*, and Δ*gltA-ppc*, showed large ATP increases ranging from ~700- to 23,000-fold (Fig. 3C,D). Notably, ATP levels alone did not fully explain susceptibility. Both Δ*ppc* and Δ*nuoG-ppc* displayed relatively modest ATP increases (~90- and ~220-fold, respectively), similar to the more resistant mutants, yet remained highly susceptible to AMP (<1% survival) (Fig. 3C,D). Thus, average ATP levels did not show a simple linear relationship with AMP-mediated killing.

Given the extensive metabolic perturbations observed in these strains, we next investigated whether AMP susceptibility was associated with altered redox balance. NADH/$NAD^+$ and NADPH/$NADP^+$ ratios were measured in WT and Δ*nuoG* (<1% survival after 60 min AMP treatment) as well as in Δ*atpA* and Δ*nuoG-gltA* (>10% survival) (Fig. 4A,B). A clear trend was observed for NADPH. The more resistant mutants, Δ*atpA* and Δ*nuoG-gltA*, exhibited only modest increases in NADPH levels (~10- to 20-fold), whereas WT and Δ*nuoG*, which were highly susceptible to AMP, showed substantially larger increases (~100- to 200-fold) (Fig. 4B). These results suggest a relationship between increased NADPH accumulation and AMP susceptibility. In contrast, NADH levels did not show a consistent association with survival. WT cells exhibited an ~500-fold increase in NADH following AMP treatment, whereas Δ*nuoG* showed only an ~10-fold increase. Similarly, Δ*atpA* displayed an ~100-fold increase, while Δ*nuoG-gltA* showed only an ~20-fold increase (Fig. 4A). Because both susceptible and resistant strains displayed either high or low NADH responses, no clear relationship between NADH levels and AMP susceptibility could be identified.

Considerable variation in both survival and metabolite levels was observed between biological replicates. To more rigorously evaluate the relationships between metabolism and AMP susceptibility, Spearman correlation analyses were performed using both initial metabolite levels and fold changes after 60 min of treatment. No significant correlations were observed between initial metabolite levels and survival (Fig. S6). In contrast, changes in ATP, formate, and NADPH levels over the course of AMP treatment all correlated significantly with survival (Spearman; ATP, $P < 0.0001$; formate, $P = 0.0005339$; NADPH, $P = 2.5 \times 10^{-6}$). NADH levels showed no significant correlation ($P = 0.2560$) (Fig. 4). Together, these results indicate that AMP susceptibility is associated with changes in ATP, formate, and NADPH metabolism, whereas NADH levels do not appear to be predictive of survival in our experimental system. Interestingly, increased NADPH accumulation correlated with decreased survival. This observation is somewhat unexpected, as elevated NADPH levels are generally associated with oxidative stress defence and protection against reactive oxygen species, suggesting that NADPH may be linked to AMP-induced metabolic remodelling through mechanisms distinct from classical oxidative stress responses.

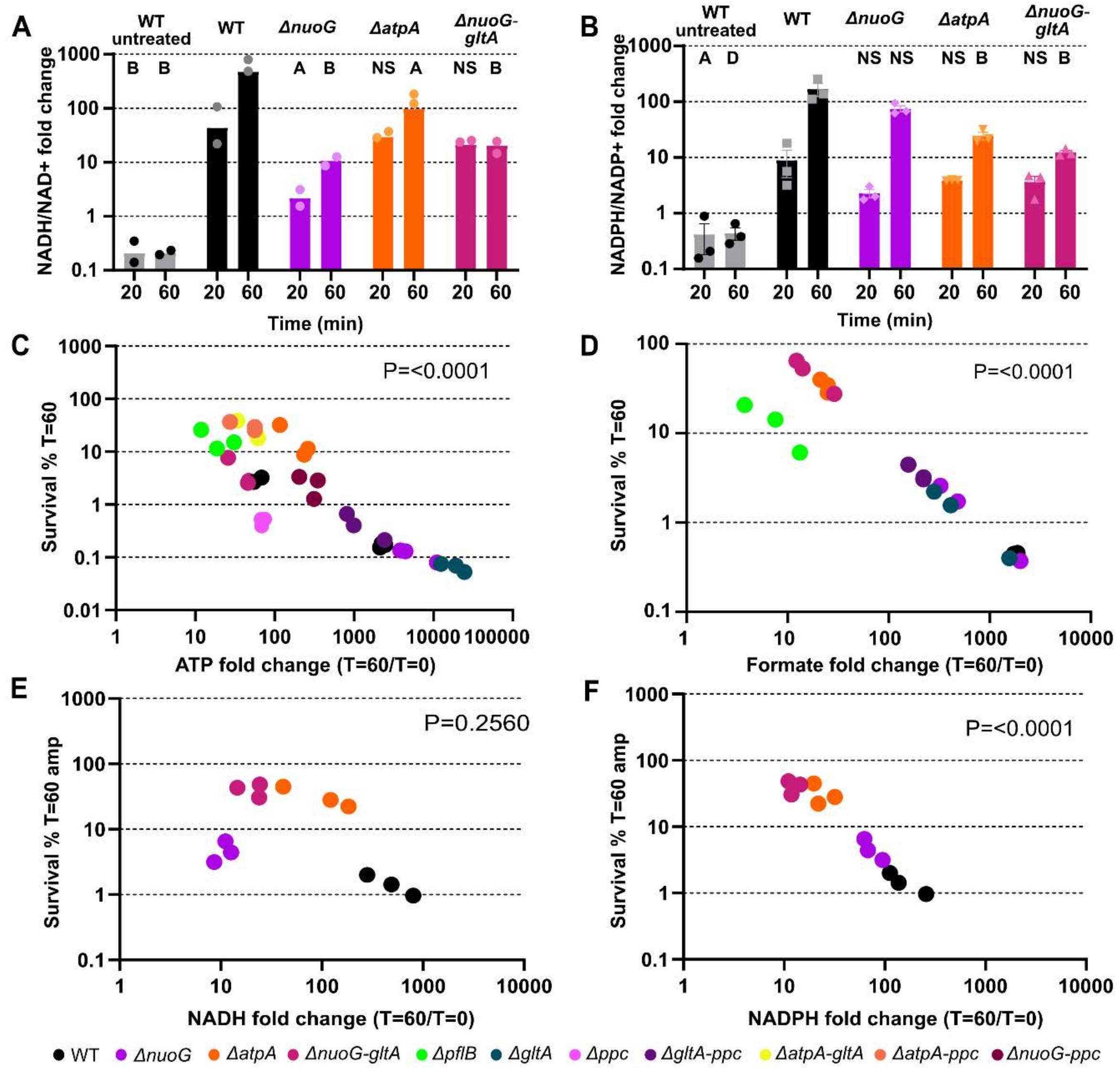

**Figure 4. Key metabolites produced in central carbon metabolism correlate with survival to AMP.** A) Fold change in NADH/NAD+ ratio. B) Fold change in NADPH/NADP+ ratio. C-F) Spearman correlation of fold-change in ATP (C), Formate (D), NADH/NAD+ (E) and NADPH/NADP+ (F) vs survival at 60 min AMP treatment. P-values are shown in respective graph.

## Discussion

How antibiotics actually kill bacteria has become a subject of intense debate, shifting from a focus on primary drug targets to the complex downstream metabolic consequences of antibiotic-target interactions[16], [17]. Understanding these metabolic responses is critical because certain mutations that block specific metabolic pathways can prevent cell death, allowing bacteria to survive doses that would otherwise be lethal [18], [19]. This matters significantly because identifying these metabolic "bottlenecks" can facilitate the development of new treatment strategies to enhance antibiotic efficacy and combat resistance.

Contrary to prevailing models that attribute antibiotic lethality primarily to reactive oxygen species (ROS) or redox imbalance [16], [20], [21], our data identify formate metabolism and ATP dynamics as central determinants of AMP killing. While formate production and secretion are closely linked to respiratory activity and cellular redox regulation via PflB[22], we find that global NAD(H)/NADP(H) levels are not reliable predictors of AMP susceptibility. Although extremely high NADH levels were associated with reduced survival, consistent with metabolic collapse, no clear correlation was observed at moderate levels. Unexpectedly, NADPH exhibited an inverse relationship with survival, with higher levels correlating with increased bacterial death, challenging the conventional view that NADPH protects against oxidative stress [23].

Instead, our results point to a model in which active metabolic flux, particularly increased formate secretion and ATP accumulation, underpins bactericidal activity. ATP, beyond its role in supporting growth, can drive detrimental processes such as dysregulated DNA replication initiation and metabolic overload [24], [25], [26]. In line with recent studies [27], our findings suggest that AMP lethality arises from energy-driven metabolic imbalance rather than redox stress, revealing a previously underappreciated link between central carbon metabolism, ATP surges, and antibiotic-induced cell death.

Surprisingly, the strongest predictor of bactericidal activity in our data is the increase in formate levels. While formate secretion itself is unlikely to be directly toxic, it appears to serve as a robust marker of the underlying metabolic rewiring that drives cell death. These findings highlight metabolic reprogramming, rather than redox stress, as a key determinant of bactericidal responses to AMP, pointing to central carbon metabolism as a potential target space for future antibiotic development. Notably, formate secretion and associated metabolic fluxes could be directly captured by our SiNWFET sensors, which resolved distinct, mutant-specific signatures driven by changes in amino acid and formate metabolism. Because these extracellular metabolic signatures correlate with bactericidal outcomes, this approach has the potential to discriminate between different modes of antibiotic action. This is particularly important given that different mutants and strains exhibit variable sensitivity to AMP, which may translate into differences in treatment efficacy. Capturing metabolic responses during antibiotic exposure therefore provides richer functional insight than a simple susceptible/resistant classification. Overall, our findings uncover key metabolic determinants of AMP killing, suggest new strategies to potentiate its efficacy, and establish SiNWFET sensors as a powerful platform for probing bacterial physiology in rapid antimicrobial susceptibility testing.

**Method:**

**Electrical test:**

All electrical data were measured by a semiconductor analyzer Keysight 4155A, cooperating with a switch unit Agilent 34970 to realize multiplexing test. Firstly, transfer curves ($I_d$ *vs.* $V_g$) of selected devices were collected in pure culture medium. Then, $I_d$ was sampled over 1 hour with a fixed $V_g$ bias in the subthreshold region. According to the measured transfer curves, $I_d$ sampling results were converted to the corresponding threshold voltage shift signals, $\Delta V_T$, which were normalized referring to the starting point. In this work, all acidification curves are averaged results from four independent devices measured simultaneously, with standard error from 3 replicates represented by shaded areas.

**Fabrication of SiNWFET sensors**

Our SiNWFETs were fabricated on silicon-on-insulator substrates using standard silicon process technology. The starting SOI wafer consisted of a 55-nm-thick lightly p-type doped silicon layer on top of 145-nm-thick buried oxide (BOX). Source and drain terminals were firstly heavily doped via arsenic ion implantation (energy: 30 keV, dose: $5\times10^{15}$ $cm^{-2}$), while the channel region was passivated by a patterned photoresist mask. After dopant activation, the device structure was defined by electron beam lithography (EBL) followed by reactive ion etching (RIE). 5-nm-thick Ni was deposited on S/D by lift-off process and subsequently annealed at 400 °C in $N_2$ for 30 s using rapid thermal process (RTP) to form NiSi. To provide an ion-sensing layer as well as surface passivation, 3-nm-thick $HfO_2$ was deposited on the chip surface via atomic layer deposition. Afterward, S/D contact pads located along the chip edges were metallized with a bilayer of 10 nm Ti and 100 nm Al. Finally, forming-gas annealing was performed in diluted $H_2$ (5% $H_2$ in $N_2$) at 400 °C for 30 min to passivate interface traps at the SiNW channel/gate-oxide interface.

**Bacterial strain and growth conditions**

*Escherichia coli* K12 MG1655 were grown directly from -80 °C freezer stocks in 5 mL LB liquid media (Sigma-Aldrich) at 37 °C in a shaking incubator at 200 rpm for 16 hours. Prior to testing all cultures were adjusted to an $OD_{600\ nm}$ of 6 in fresh media.

*E. coli* mutants were made by P1 transduction from KEIO collection strains. Overnight cultures of the KEIO collection strains were sub-cultured 1:100 in fresh LB media with $CaCl_2$ and grown at 37 °C, 200 rpm, to $OD_{600\ nm}$ 0.1 before addition of plain P1 phage lysate and continued to incubate until visible lysis had occurred. Lysed cells were incubated with chloroform for 5 mins prior to centrifugation and lysate collection. Prepared lysates were added to cultures of *E. coli* MG1655, $OD_{600\ nm}$ 0.6-1 and incubated 37 °C for 30 mins. NaCitrate was added prior to resuspension in LB with NaCitrate and incubated for 2 h, 37 °C. Cells were collected and plated onto LB agar containing kanamycin and grown at 37 °C overnight.

Double mutants were made by initially removing one gene using the phage lysate protocol and removing the kanamycin cassette by FLP recombinase expressed from pCP20. This allowed for phage transduction of the second gene.

All mutants were confirmed by PCR amplification of the gene of interest. All primers are listed in supplementary table S4.

**Metabolic profiling and electrical measurement (on-chip protocol)**

At time of testing, the fabricated chip and polydimethylsiloxane (PDMS) container were connected to the measurement setup via a probe card on a 37 °C preheated stage. An initial 50 µL of test media with 20 µL fluorinert FC-70 oil was added in the PDMS container with an electrode placed inside. Transfer curves were used to determine a stable baseline. *E. coli* cultures were added with or without AMP and additional fluorinert FC-70 oil was added. $I_d$ vs. *time* measurement was continued for a minimum of 50 min. Detailed procedures for the on-chip assay are provided in the SI. In these sections, we systematically evaluated key experimental factors, including the injection method, temperature control (Fig. S7), the buffering capacity of LB medium (Fig. S8), and glucose supplementation (Fig. S9).

**Time killing optical density**

Overnight cultures of *E. coli* were resuspended with fresh LB broth containing 10 g/L glucose and incubated with or without AMP at 37 ˚C. Measurements were taken using a benchtop Spectrophotometer (WPA biowave CO8000) every 10 min for 1 h.

**Time killing on solid media**

Overnight cultures of *E. coli* were resuspended with fresh LB broth containing 10 g/L glucose and setup up in a PDMS ring on a silicon chip to replicate electrical data sampling and incubated at 37 °C with or without AMP for 2 h with 20 µL samples taken every 20 min. The samples were serially diluted in PBS and plated onto LB agar and incubated overnight at 37 °C. The individual colonies were counted and survival ratio calculated.

**Preparation of lysed cell suspension**

A 20 mL overnight culture of *E. coli* was resuspended in 1 mL of LB medium supplemented with 10 g/L glucose and a sample taken to determine initial CFU/mL. The culture was transferred to a microtube containing 0.5 g of glass beads and subjected to mechanical cell lysis using a BeadBug6 microtube homogenizer. Lysis was performed for 20 cycles at 4350 rpm, with 20-second rest intervals between cycles. Following homogenization, the lysate was collected, a sample taken to determine CFU/mL and stored at 4 °C until use in subsequent experiments. The cell density and lysis percentage can be found in Fig. S10.

**ATP detection**

ATP detection was conducted using Promega BactitreGlo. *E. coli* with or without AMP were prepared following the on-chip protocol and BactitreGlo reagents were prepared according to manufacture instructions. Samples were taken at required time points for CFU/mL determination and ATP detection. The samples for ATP were added to a white bottom 96-well plate, incubated at RT for 5 mins with 200 rpm prior to luminescence read (TECAN infinite M200 pro). Luminescence values were interpolated against a standard curve for nM values. Samples taken for CFU/mL determination were incubated 37 °C overnight, colonies counted and used for determination of ATP per cell by correcting against the nM value.

**Method of LC-MS**

The detailed method of LC-MS can be found in SI.

**NADH/NAD+ and NADPH/NADP+ detection**

NADH/NAD+ And NADPH/NADP+ detection was conducted using Promega Glo-Assay kits. *E. coli* with or without AMP were prepared following the on-chip protocol and BactitreGlo reagents were prepared according to manufacture instructions. Samples were taken at required time points for CFU/mL determination and NADH/NAD+ and NADPH/NADP+ detection. The samples were added to a white bottom 96-well plate, incubated at RT for 30 mins at room temp and luminescence read (TECAN infinite M200 pro). Luminescence values were interpolated against a standard curve for nM values. Samples taken for CFU/mL determination were incubated 37 °C overnight, colonies counted and used for determination of NADH/NAD+ or NADPH/NADP+ per cell by correcting against the nM value.

**Formate detection**

Formate detection was conducted using a formate assay kit from Sigma-Aldrich. *E. coli* with or without AMP were prepared following the on-chip protocol and formate kit reagents were prepared according to manufacture instructions. Samples were taken at required time points for CFU/mL determination and formate detection. The samples were added to a clear bottom 96-well plate, incubated at RT for 60 mins at room temp and $OD_{565\ nm}$ recorded (TECAN infinite M200 pro). Formate concentrations were determined by interpolation against a standard curve. Samples taken for CFU/mL determination were incubated 37 °C overnight, colonies counted and used for determination of formate per cell by correcting against the nM value.

**Data analysis**

All the electrical measurement data were collected by the HP4155 Semiconductor Parameter Analyzer controlled by Labview software. The $I_{DS}$ curve was converted to $V_T$ shift ($\Delta V_T$) based on the transfer curve by the program running on Python. All other data was compiled and processed in Excel and data analysis conducted using GraphPad Prism 10. All t test data analysis was conducted using GraphPad Prism 10, the data was split into 20 min and 60 min time points for comparison with all data compared to *E. coli* WT AMP treated. The electrical data (Fig 3B) used an unpaired, normal (Gaussian) Welch's t test. All other data used an unpaired, lognormal, Welch's t test.

# Supplementary Materials for

## Nanoscale silicon sensor – guided new insights into early metabolic response of *Escherichia coli* to ampicillin

Yingtao Yu[†], Victoria Nolan[†], Zheqiang Xu, Allison Jones, George Alhoush, Zhen Zhang*, Sanna Koskiniemi*

*Corresponding author: Zhen Zhang, zhen.zhang@angstrom.uu.se
Sanna Koskiniemi, sanna.koskiniemi@icm.uu.se;

**This PDF file includes:**

Supplementary Text
Figs. S1 to S10
Tables S1 to S4

# Supplementary Text

## 1. On-chip test procedure

During on-chip metabolism monitoring, the bacterial suspension was initiated at the stationary phase to rule out the signal contribution from bacterial growth, with fresh LB broth. After *E. coli* overnight culturing, 200 μL of bacterial suspension was first centrifuged at 4500 rpm. After removing the supernatant, the precipitate was resuspended in fresh LB in either 40 μL (test1) or 133 μL (test 2) fresh LB. At the beginning of the measurement, either 30 μL (test 1) or 100 μL (test 2) of the resuspended bacteria was pipette-injected into the container where background LB ($V_3$) was already placed for baseline settlement. The final bacterial concentration in the container ($V_2$+$V_3$) was adjusted to saturated value by manipulating the dilution ratio in this cell loading process. Since our stage cannot be shaken during testing, it is essential to ensure effective mixing of the suspension during the injection step. Here, two different initial conditions were tested with alterations to the volumes of bacterial suspension along with their repeated test results presented in Fig. S8A.

The incubation curves with the same concentration of bacteria in the container were dependent on the injection step as shown in Fig. S8A. For a higher injection concentration of bacteria, such as Test-1, the measured $\Delta V_T$ increased more rapidly, which coincides with the previous work [1]. However, the sharp signal drop was often observed in repeated measurements. When the bacterial sedimentation was resuspended with 133 μL LB broth in Test-2, the slope of $\Delta V_T$ *vs.* time decreased but reached the same level as Test-1 after 20 minutes, with excellent repeatability. The only difference between the two tests was the initial cell density of the injected bacteria. When the high-density bacterial suspension (Test-1) was injected, over-condensed bacteria could settle to the surface of the sensor bottom, where cells could attach to the SiNW channel leading to a locally higher signal response. Additionally, the temperature gradient in the container could also cause the signal dependence on location. To verify these hypotheses, both Test-1 and Test-2 were repeated with a filter paper of 0.45 μm pore size embedded in the container to isolate cells from the channel surface. As shown in Fig. S8B, the acidification rate gap still existed between the two tests, excluding the contribution from cell attachment, suggesting that this is not the cause of the difference between the two tests, and all further experiments were conducted without the filter paper. Different incubation temperatures on the chip were also verified, rising from 34 to 40 °C within the tolerance range for *E. coli*, as presented in Fig. S8C. Obviously, the acidification was accelerated at a higher medium temperature. A plausible explanation is that increased kinetic energy of cellular components at higher temperatures could lead to an elevated overall metabolic rate [2], increasing the acetate overflow reactions. Despite the observed temperature dependence, rapid acidification of Test-1 did not appear even at 40 °C, implying that increased incubation cannot sufficiently boost the acidification process. As possible objective factors were ruled out, the variation of acidification curves in Test-1 most probably originated from the bacteria themselves due to a higher initial inoculum. In our previous work, it was observed that the $H^+$ ion production rate per cell increases with cell density, which was ascribed to the metabolic adjuvant role of dead cells (cell growth and death could both take place when the cell density is high) [1]. Therefore, the metabolic rate of the locally condensed bacteria in Test-1 could also be accelerated for the same reason, resulting in the measured rapid acidification. Considering the repeatability and controllability, all other tests in this manuscript strictly followed the procedure of Test-2 at 37 °C.

## 2. Culture media for on-chip test

To verify the buffering capacity of LB medium, we titrated LB and, for comparison, a 0.9% NaCl aqueous solution with a series of HCl concentrations (Fig. S9A). A near-Nernstian pH response was observed in 0.9% NaCl, indicating a negligible buffering effect. In contrast, the $V_T$ response in LB was strongly suppressed, reaching only ~30 mV even after adding an amount of HCl equivalent to lowering the solution to pH 3. These results clearly demonstrate that LB provides substantially stronger buffering than the 0.9% NaCl solution.

Time-kill assays were then performed in both media using $4 \times 10^9$ CFU/mL wild-type *E. coli* (Fig. S9B). After 1 h exposure to 100 µg/mL ampicillin, the survival fraction in LB decreased to below 1%. By comparison, the survival fraction remained close to 100% in 0.9% NaCl under the same AMP treatment, indicating no killing effect. This discrepancy is likely due to the rapid acidification of the unbuffered 0.9% NaCl solution, which suppresses cellular metabolism and thereby diminishes the killing efficacy of AMP.

## 3. Concentrations of AMP and glucose

The extracellular pH change is also dependent on the carbon source participating in the metabolic pathway. For instance, extracellular pH decreases if glucose, glycerol, or octanoate are used as carbon sources. However, the growth media becomes alkalinized with more oxidized carbon sources, such as citrate, 2-furoate, and 2-oxoglutarate [3]. During our measurements, LB broth with glucose was utilized as the culture medium, containing glucose, and mixed peptides from tryptone and yeast extract as carbon sources. The carbon sources are continuously consumed during tests, which might alter the metabolic activity and pathways. To identify the signal contribution from carbon sources, a series of tests was conducted with varying initial glucose concentrations, as shown in Fig. S10A. In glucose-free LB medium, $\Delta V_T$ experiences a slight increase in the initial 10 min, followed by a continuous decrease. This is likely due to the strict preferential order of amino acid consumption in *E. coli* [4]. First, serine and aspartate are consumed with concomitant acetate formation, detected as $\Delta V_T$ increases. Then, tryptophan is utilized accompanied by acetate uptake, resulting in a plateau and subsequent decrease (see purple line in Fig. S10A). However, LB medium with glucose shows a constant acidification rate regardless of initial glucose concentration (Fig. S10A). Surprisingly, even with the lowest glucose concentration (0.1 g/L), the acidification rate, as indicated by $\Delta V_T$ slope, initially follows the same rate of higher glucose concentrations before the glucose was consumed at 12 min. Then, the observed medium alkalization might suggest that amino acids are utilized as a carbon source for metabolism. To ensure sufficient glucose availability, LB medium supplemented with 10 g/L glucose was used in all experiments described in this manuscript. The same concentrations of glucose were then used alongside AMP treatment of *E. coli* (Fig. S10B). The lowest concentration of glucose, 0.1 g/L, acidification curve shows a decrease after 15 min as the glucose was depleted, which is consistent with the result of the untreated sample in Fig. S10B. However, all other glucose concentrations exhibit the same decrease with an onset time of 20 minutes under AMP treatment, regardless of glucose concentration. Therefore, the influence of glucose concentration can largely be discounted when observing the metabolic response to antibiotic treatment.

## 4. Cell lysis

To estimate the medium pH change caused by cell lysis under AMP treatment, we artificially lysed bacterial suspensions using a bead-beating protocol. In this process, beads are mixed with the cell suspension, and mechanical agitation disrupts the bacterial cells. To assess lysis efficiency, overnight-cultured WT *E. coli* samples (3 biological replicates) were subjected to bead beating (Fig. S11A). After 20 cycles, the viable cell density decreased from ~$1 \times 10^9$ to ~$1 \times 10^7$ CFU/mL. After centrifugation, the viable count was further reduced to ~$1 \times 10^6$ CFU/mL. The corresponding lysis fraction is summarized in Fig. S11B. After 20 cycles, the lysis fraction reached 98.5%, indicating that the bead-beating protocol provides sufficiently high lysis efficiency for this estimation.

In Fig. S2B, to mimic a fully lysed condition, 20 mL of overnight-cultured WT *E. coli* was concentrated to 1 mL and subjected to 20 cycles of bead-beating lysis. After 1 h of on-chip incubation with a 150 µL sample, 7.5 µL of the lysis supernatant, corresponding to the supernatant from a fully lysed 150 µL sample, was added to the container, resulting in a negligible change in $V_T$. Additional lysis supernatant was then added sequentially (up to 4 × 7.5 µL), yet no significant $V_T$ decrease was observed. These results demonstrate that cell lysis alone does not induce the pronounced $V$t drop observed under AMP treatment.

## 5. Method of LC-MS

The chromatographic separation was performed on an Agilent 1290 Infinity UHPLC system (Agilent Technologies, Waldbronn, Germany). 1 µL of each sample was analyzed on a Luna Omega Polar C18, 100 × 2.1 mm, 1.6 µm in combination with a 2.1 mm SecurityGuard cartridge (Phenomenex, Torrance, CA, USA) by using gradient elution of 0.1% formic acid (v/v) in water as mobile phase A and acetonitrile/isopropanol (70/30, v/v) as mobile phase B.

The mobile phase was delivered on the column by a flow rate of 0.4 mL/min with the following gradient: 0-1 min (5% B), 5 min (30% B), 9 min (50% B), 12 min (78% B), 15 min (95% B), 16 min (100% B), 18 min (100% B), 18.1 min (5% B), 20 min (5% B). The column and autosampler were thermostated at 40 °C and 4 °C, respectively. The compounds were detected with an Agilent 6546 Q-TOF mass spectrometer equipped with a jet stream electrospray ion source operating in positive or negative ion mode. The settings were kept identical between the modes, with the exception of the capillary voltage. A reference interface was connected for accurate mass measurements; the reference ions purine (4 µM) and HP-0921 (Hexakis(1H, 1H, 3H-tetrafluoropropoxy)phosphazine) (1 µM) were infused directly into the MS at a flow rate of 0.05 mL/min for internal calibration, and the monitored ions were purine m/z 121.05 and m/z 119.03632; HP-0921 m/z 922.0098 and m/z 966.000725 for positive and negative mode respectively. The gas temperature was set to 150°C, the drying gas flow to 8 L/min, and the nebulizer pressure 35 psi. The sheath gas temperature was set to 350 °C, and the sheath gas flow was 11 L/min. The capillary voltage was set to 4000 V in positive ion mode and to 4000 V in negative ion mode. The nozzle voltage was 300 V. The fragmentor voltage was 120 V, the skimmer 65 V, and the OCT 1 RF Vpp 750 V. The collision energy was set to 0 V. The m/z range was 50 - 1000, and data were collected in centroid mode with an acquisition rate of 4 scans/s (2265 transients/spectrum).

# Supplementary information

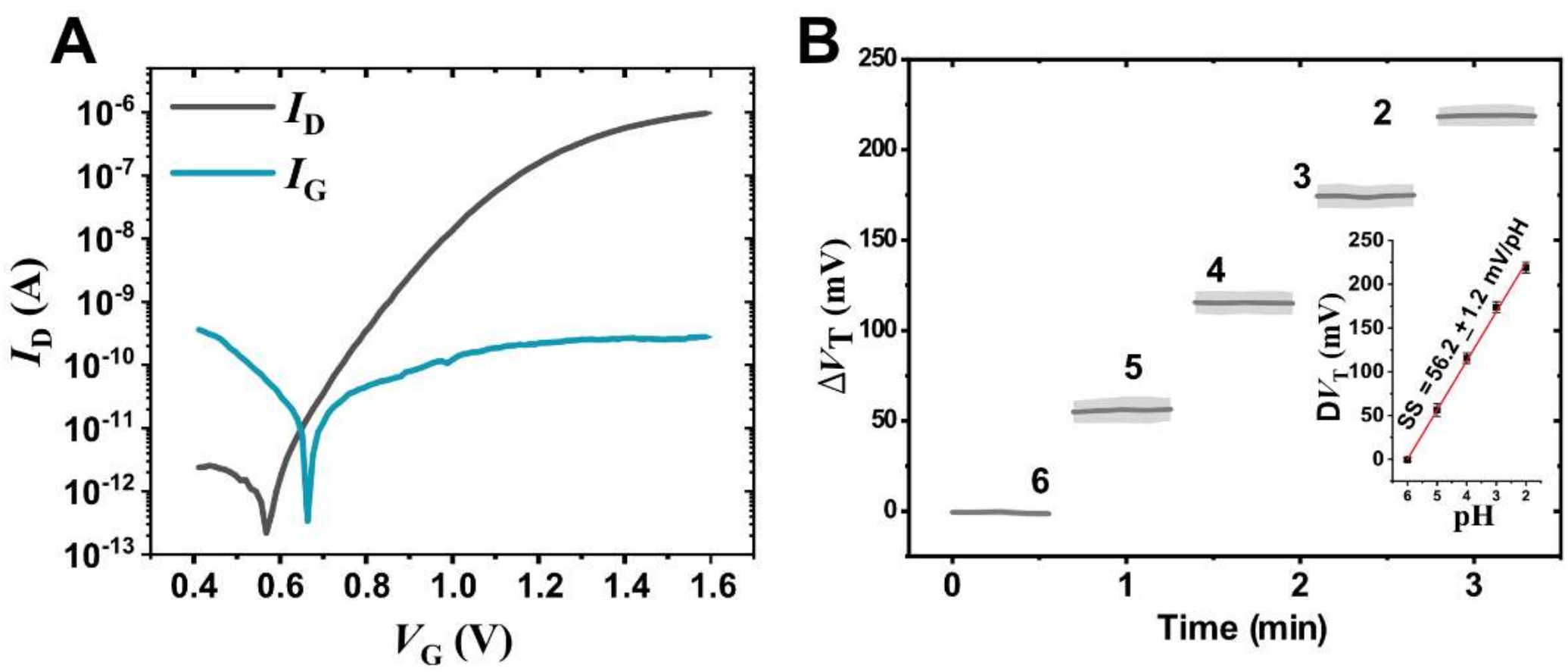

Figure S1. A) Transfer curve $I_D$ vs. $V_G$ of SiNWFET measured in LB under 0.3 V $V_D$ , and B) $\Delta V_T$ vs. $t$ averaged from 3 SiNWFETs measured in solutions with *p*H changing from 6 to 2. The pH values were annotated on the curve. Inset: SiNWFET sensitivity calibration curve, *i.e.*, $\Delta V_T$ vs. pH.

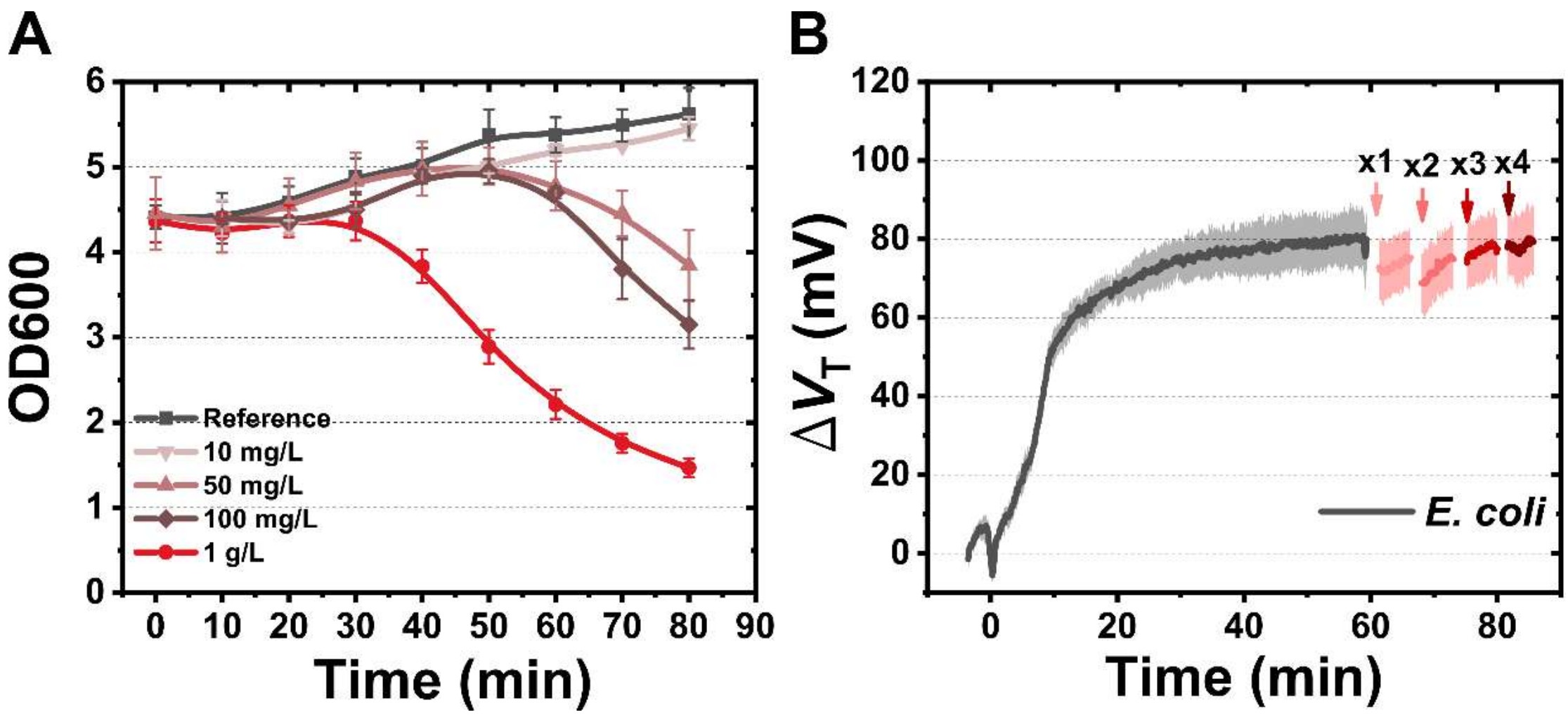

Figure S2. A) $OD_{600}$ of bacterial suspension versus time under different AMP treatment conditions. B) $\Delta V_T$ versus time plot with multiple injected lysed WT *E. coli* suspensions after 1-h incubation showing no significant signal drop.

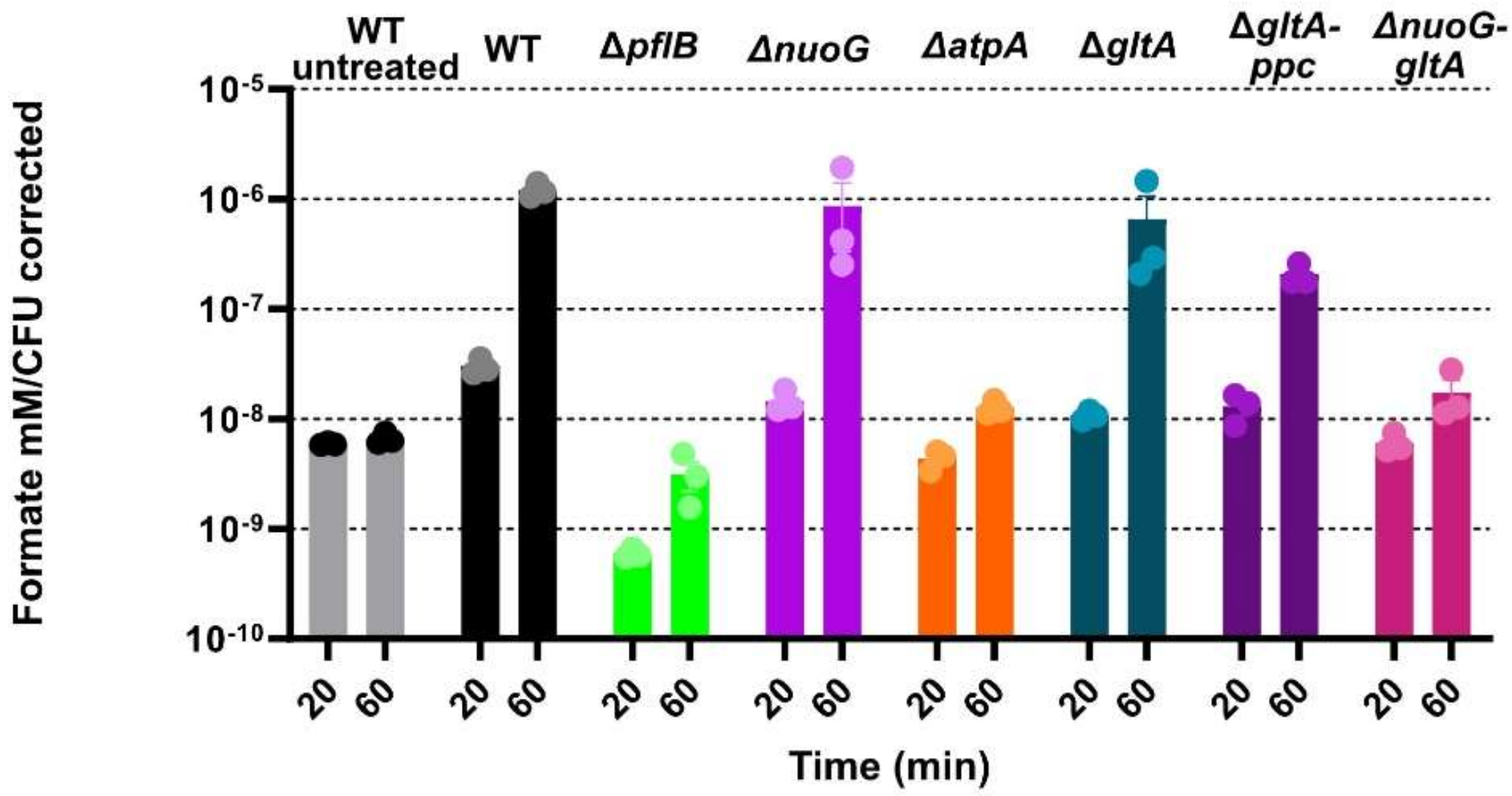

Figure S3. Formate nM/CFU concentrations for *E. coli* WT and select *E. coli* metabolic mutants.

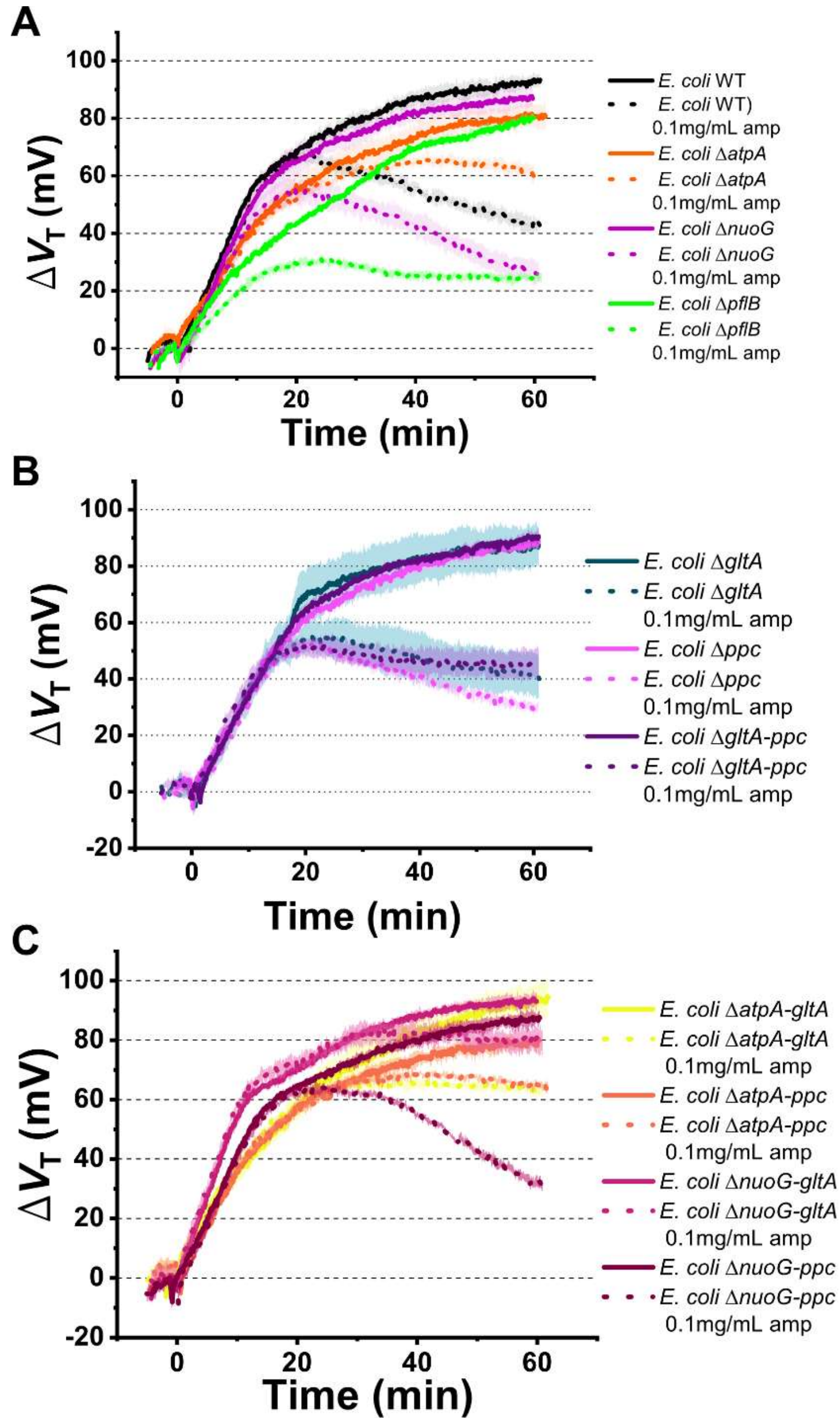

Figure S4. $\Delta V_T$ versus time plot of untreated (solid lines) and 100 mg/L AMP-treated (dashed lines) *E. coli* mutants, from which we extracted the $\Delta V_T$ histogram, Fig. 3B.

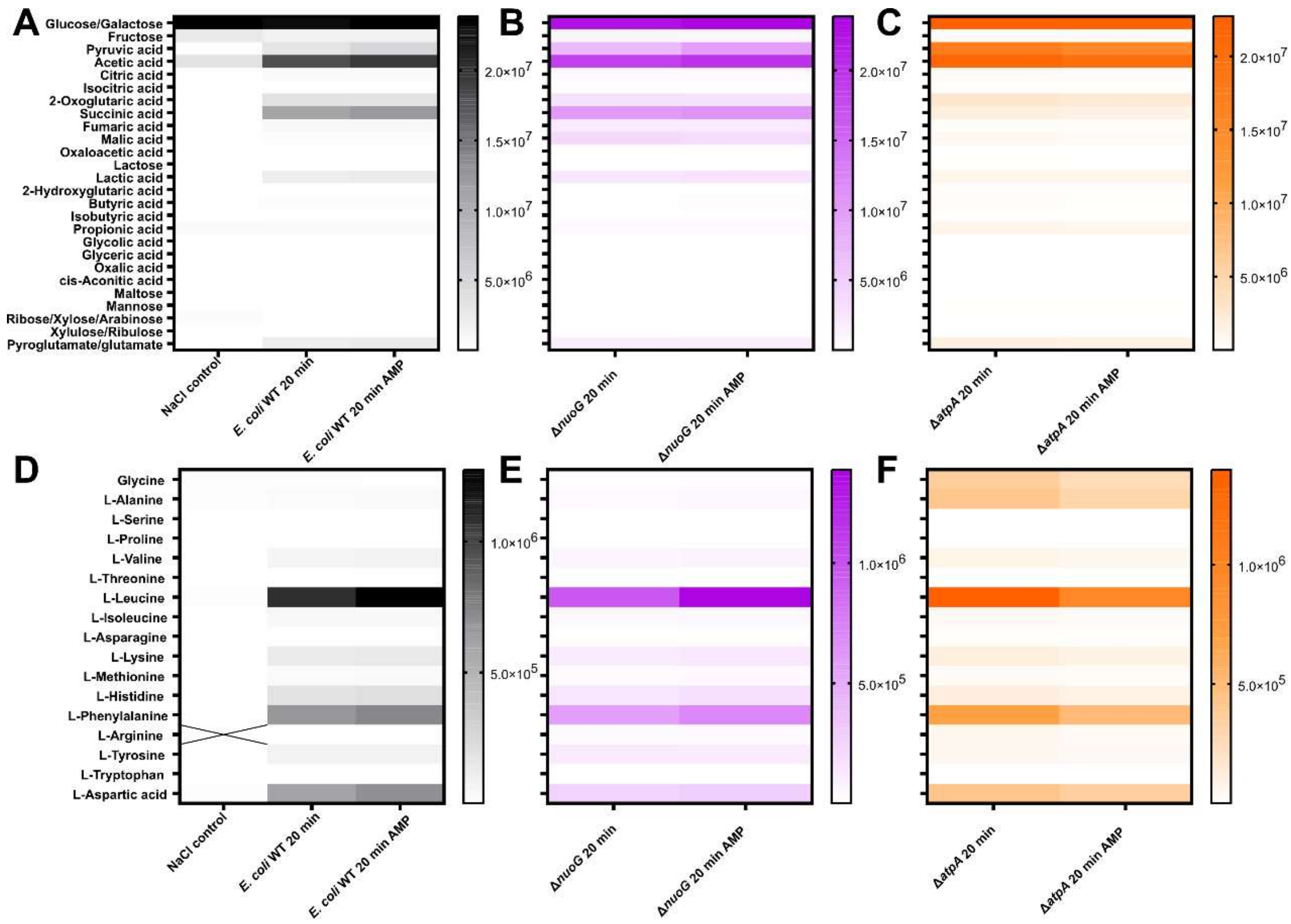

Figure S5. Metabolomics analysis of *E. coli* in 0.9 % NaCl, 1 % glucose media. A, B and C) Organic acids secreted by *E. coli* WT, Δ*nuoG* and Δ*atpA* with and without AMP treatment. Amino acids secreted by *E. coli* WT, Δ*nuoG* and Δ*atpA* with and without AMP treatment.

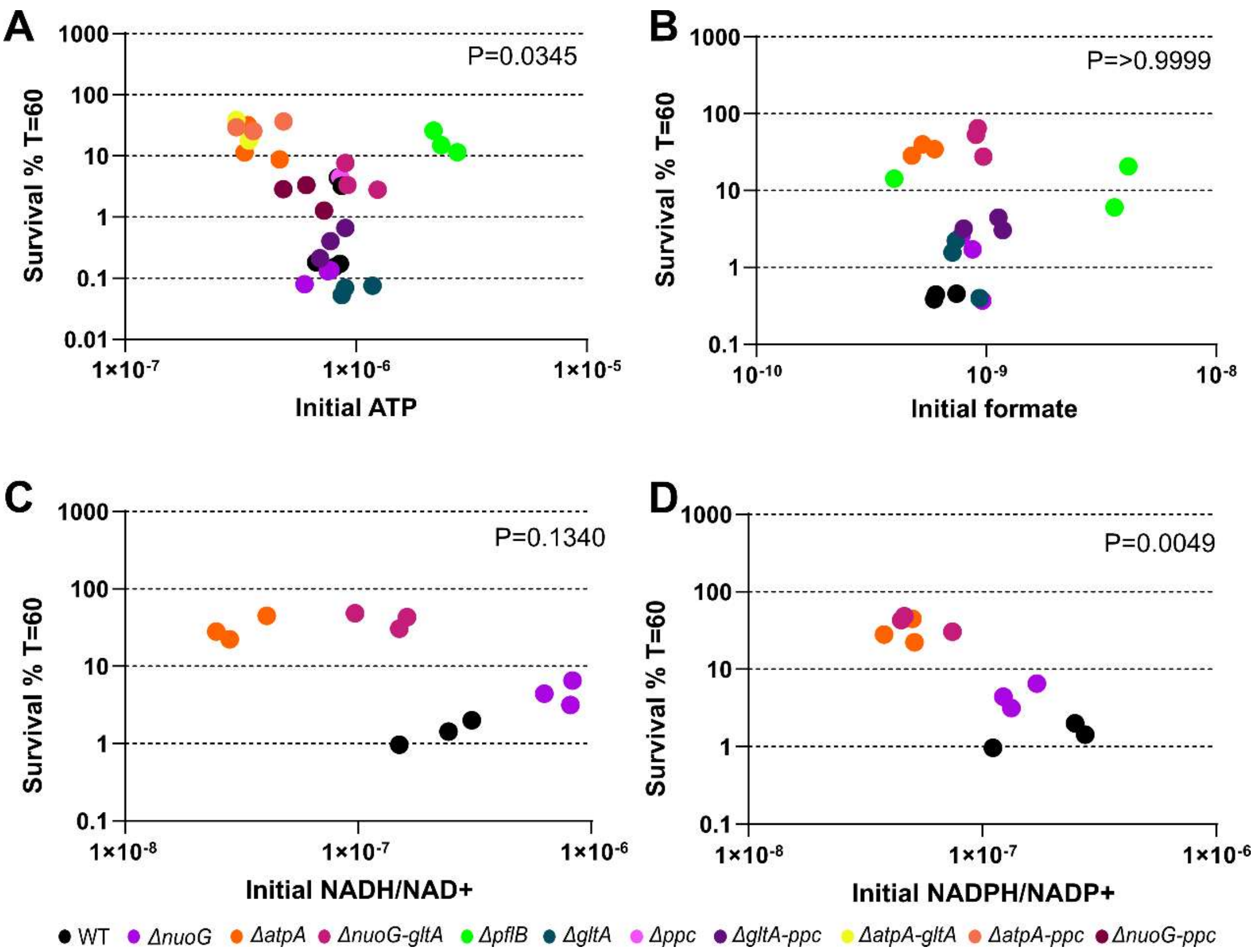

Figure S6. Correlation scatter plots of initial metabolite concentrations vs survival % against AMP by 60 min. A) ATP B) Formate C) NADH/NAD+ D) NADPH/NADP+.

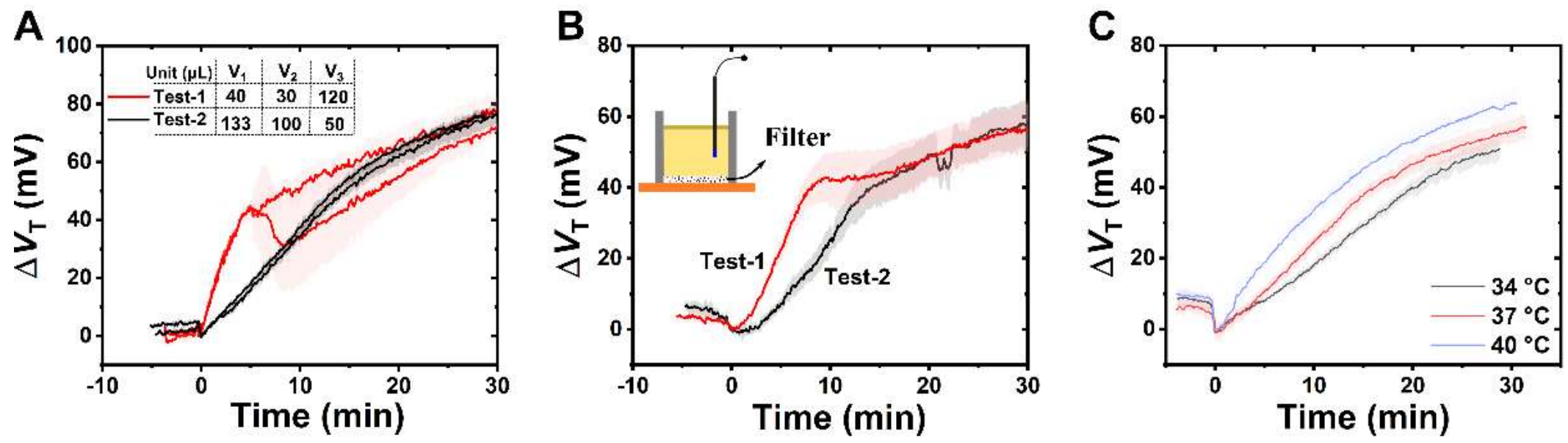

Figure S7. *E. coli* acidification curves, $\Delta V_{\mathrm{T}}$ *vs*. *time* A) for repeated tests with different dilution processes at 37 °C, B) with a filter to isolate bacteria from the chip surface, and C) at different media temperatures following Test-2.

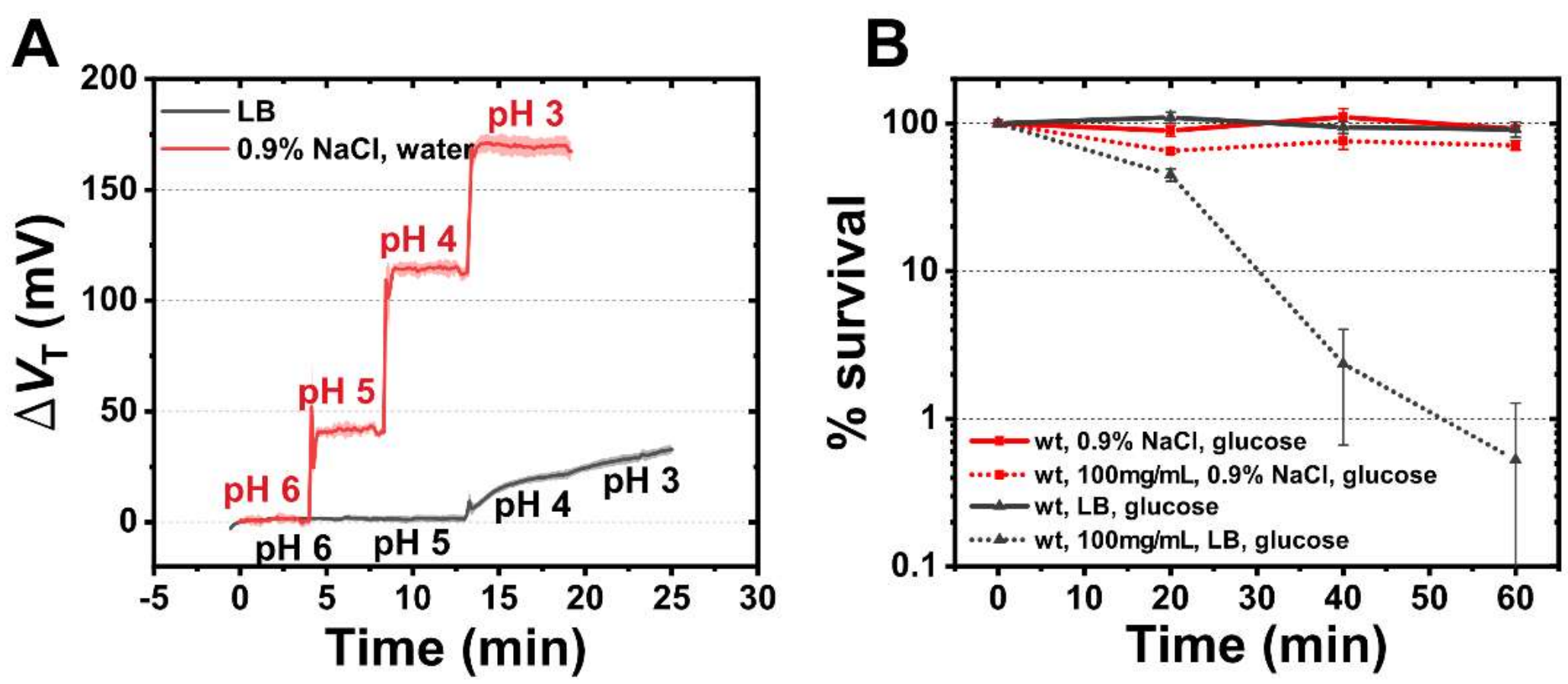

Figure S8. A) $\Delta V_T$ *vs*. *time* of pH adjustment by adding HCl acid into 0.9% NaCl water solution and LB. B) Time-killing results of $4 \times 10^9$ CFU/mL wt *E.coli* in both 0.9% NaCl water solution and LB.

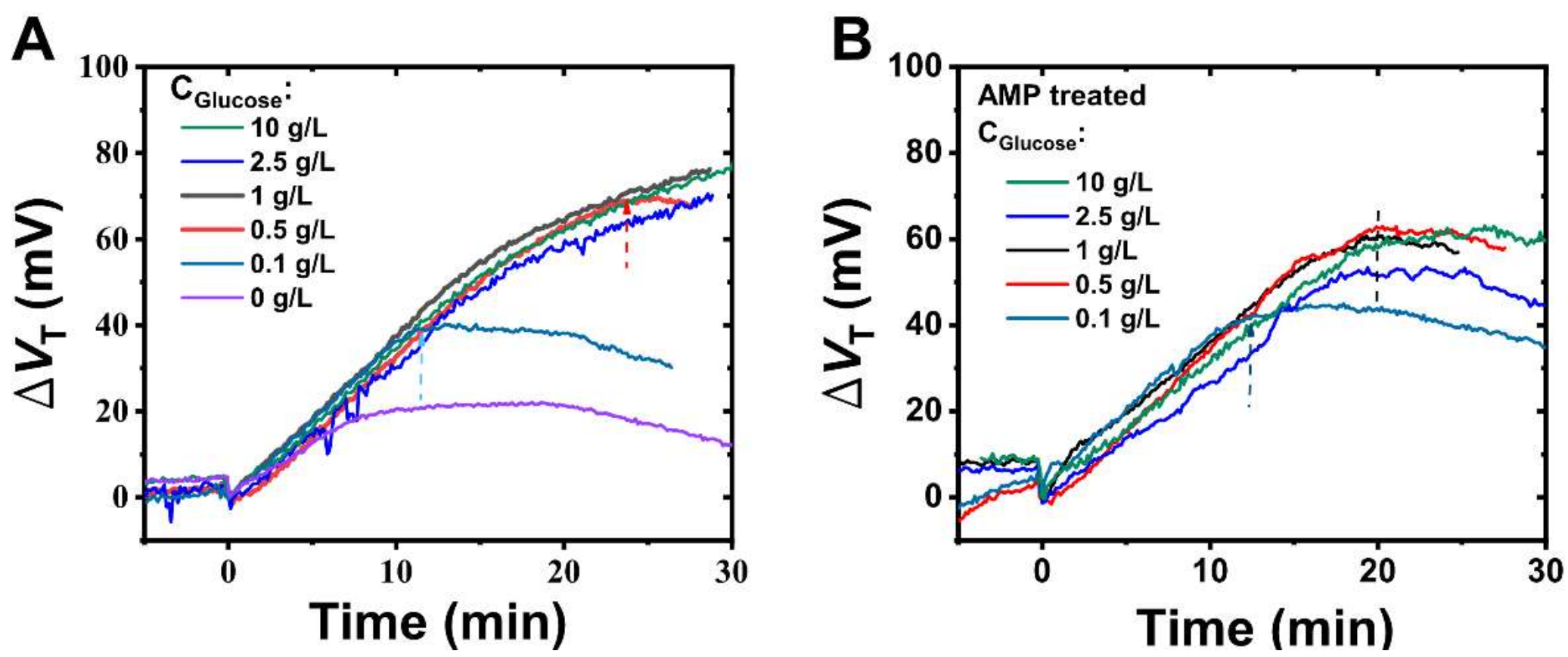

Figure S9. $\Delta V_T$ *vs*. *time* curves for *E. coli* with different glucose concentrations A) without AMP, B) with AMP (100 mg/L) treatment.

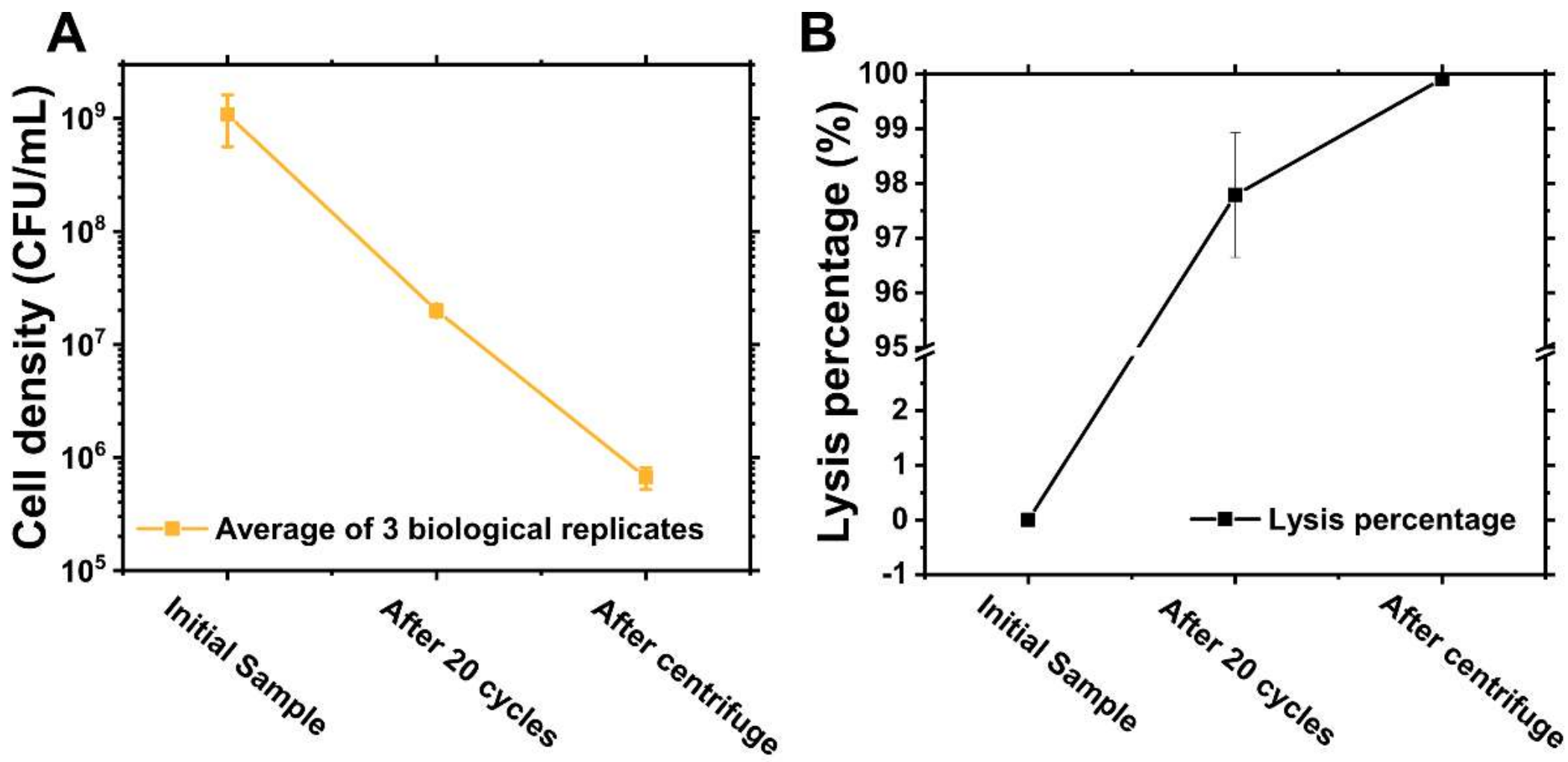

Figure S10. A) cell density for WT *E. coli* after bead beating lysis cycles and centrifugation. B) Calculated lysis percentage based on data from A).

**Table S1**. Welch's t-test P values comparing all strains to *E. coli* WT 100 µg/mL AMP at 20 mins.

| Strain | Electrical data | Survival | ATP | Formate | NADH/NAD+ | NADPH/NADP+ |
|---|---|---|---|---|---|---|
| WT | 0.0007 | <0.0001 | 0.0165 | 0.0325 | 0.016 | 0.0138 |
| Δ*pflB* | <0.0001 | <0.0001 | 0.0208 | 0.0235 | | |
| Δ*nuoG* | 0.0036 | 0.0140 | 0.4737 | 0.0454 | 0.0125 | 0.1460 |
| Δ*atpA* | <0.0001 | <0.0001 | 0.3694 | 0.0298 | 0.4933 | 0.3844 |
| Δ*gltA* | 0.0144 | 0.3700 | 0.0351 | 0.0416 | | |
| Δ*ppc* | <0.0001 | 0.4924 | 0.2376 | | | |
| Δ*gltA-ppc* | 0.0004 | 0.1527 | 0.0300 | 0.0375 | | |
| Δ*atpA-gltA* | 0.0018 | <0.0001 | 0.2172 | | | |
| Δ*atpA-ppc* | 0.0010 | <0.0001 | 0.1101 | | | |
| Δ*nuoG-gltA* | 0.0434 | <0.0001 | 0.0196 | 0.0285 | 0.2585 | 0.3115 |
| Δ*nuoG-ppc* | 0.0140 | 0.0407 | 0.6243 | | | |

**Table S2**. Welch's t-test P values comparing all strains to *E. coli* WT 100 µg/mL AMP at 60 mins

| Strain | Electrical data | Survival | ATP | Formate | NADH/NAD+ | NADPH/NADP+ |
|---|---|---|---|---|---|---|
| WT | 0.0004 | <0.0001 | <0.0001 | 0.0012 | 0.0011 | <0.0001 |
| Δ*pflB* | <0.0001 | <0.0001 | 0.0027 | 0.0012 | | |
| Δ*nuoG* | 0.0020 | 0.4397 | 0.1044 | 0.2284 | 0.0027 | 0.0737 |
| Δ*atpA* | <0.0001 | <0.0001 | 0.0090 | 0.0012 | 0.0485 | 0.0058 |
| Δ*gltA* | 0.7447 | 0.0079 | 0.0067 | 0.1100 | | |
| Δ*ppc* | 0.0002 | 0.4299 | 0.2181 | | | |
| Δ*gltA-ppc* | 0.3355 | 0.7906 | <0.0001 | 0.0006 | | |
| Δ*atpA-gltA* | <0.0001 | <0.0001 | 0.0016 | | | |
| Δ*atpA-ppc* | <0.0001 | <0.0001 | 0.0029 | | | |
| Δ*nuoG-gltA* | <0.0001 | <0.0001 | 0.0068 | 0.0011 | 0.0024 | 0.0056 |
| Δ*nuoG-ppc* | 0.0005 | 0.5886 | 0.0039 | | | |

**Table S3**. *E. coli* mutant strains utilized for exploring metabolic changes in response to AMP

| Strain | Gene | Consequence |
|---|---|---|
| *E. coli Δicd* | isocitrate dehydrogenase enzyme | Prevents the formation of α-ketoglutarate from isocitratric acid and forces the TCA cycle to continue down the glyoxylate shunt |
| *E. coli ΔgltA* | citrate synthase enzyme | Unable to produce citric acid and proceed into the full TCA cycle |
| *E. coli Δppc* | phosphoenolpyruvate carboxylase | Cannot utilise the conversion of phosphoenolpyruvate into oxaloacetic acid for increased TCA cycle activity between oxaloacetic acid and succinate |
| *E. coli ΔgltA-ppc* | citrate synthase enzyme and phosphoenolpyruvate carboxylase | Unable to use the TCA cycle completely |
| *E. coli ΔatpA-gltA* | α subunit of F1 ATP synthase and citrate synthase enzyme | Non-function ATP synthase complex preventing ATP generation from the ETC and unable to produce citric acid and proceed into the full TCA cycle |
| *E. coli ΔatpA-ppc* | α subunit of F1 ATP synthase and phosphoenolpyruvate carboxylase | Non-functioning ATP synthase complex preventing ATP generation from the ETC and cannot utilise the conversion of phosphoenolpyruvate into oxaloacetic acid for increased TCA cycle activity between oxaloacetic acid and succinate |
| *E. coli ΔnuoG-gltA* | the subunit G of NADH dehydrogenase I and citrate synthase enzyme | Disruption to the first step of the ETC and unable to produce citric acid and proceed into the full TCA cycle |
| *E. coli ΔnuoG-ppc* | the subunit G of NADH dehydrogenase I and phosphoenolpyruvate carboxylase | Disruption to the first step of the ETC and cannot utilise the conversion of phosphoenolpyruvate into oxaloacetic acid for increased TCA cycle activity between oxaloacetic acid and succinate |

**Table S4.** Primers used for PCR confirmation of mutant strains

| Strain | Primer |
|---|---|
| atpA | AAGTCTGTAATGGCAGGCG – F<br>TTTGCGTGTTCTGGACGC - R |
| nuoG | GTGGAGCCGTTACAGAGC – F<br>AGCAGGATCTCAATCAGTTCC - R |
| gltA | ATCAACCCGCCATATGAACG – F<br>AAGCCAGGTTGATGTGCGAA – R |

| ppc | CGAGGGTGTTAGAACAGAAG – F<br>CGATTTCGCAGCATTTGACG – R |
|---|---|